\documentclass[apjl]{emulateapj}

\lefthead{Cassano et al.}
\righthead{On the connection between radio halo \& cluster mergers}

\def\ltapprox{\raise 2pt \hbox {$<$} \kern-1.1em \lower 5pt \hbox {$\approx$}}
\def\ltsim{\raise 2pt \hbox {$<$} \kern-1.1em \lower 4pt \hbox {$\sim$}}
\def\gtsim{\raise 2pt \hbox {$>$} \kern-1.1em \lower 4pt \hbox {$\sim$}}

\def\ie{{\it i.e.,~}}
\def\eg{{\it e.g.,~}}
\def\gtsim{\; \raise0.3ex\hbox{$>$\kern-0.75em \raise-1.1ex\hbox{$\sim$}}\; }
\def\ltsim{\; \raise0.3ex\hbox{$<$\kern-0.75em \raise-1.1ex\hbox{$\sim$}}\; }

\shorttitle{On the connection between giant radio halos and cluster mergers}
\shortauthors{Cassano et al.}

\begin{document}

\title{On the connection between giant radio halos and cluster mergers}

\author{R. Cassano\altaffilmark{1,2}
S. Ettori\altaffilmark{3}, 
S. Giacintucci\altaffilmark{1,2},
G. Brunetti\altaffilmark{1},
M. Markevitch\altaffilmark{2},
T. Venturi\altaffilmark{1},
M. Gitti\altaffilmark{3}}

\altaffiltext{1}{INAF/IRA, via Gobetti 101, I--40129 Bologna, Italy}
\altaffiltext{2}{Harvard-Smithsonian Center for Astrophysics, 60 Garden Street, Cambridge, MA 02138}
\altaffiltext{3}{INAF/Osservatorio Astronomico di Bologna, via Ranzani 1, I--40127 Bologna, Italy}


\begin{abstract}

The frequently observed association between giant radio halos and merging galaxy clusters has driven present theoretical models of non-thermal emission from galaxy clusters, which are based on the idea that the energy dissipated during cluster-cluster mergers could power the formation of radio halos. To quantitatively test the merger-halo connection we present the first statistical study based on deep radio data and X-ray observations of a complete X-ray selected sample of galaxy clusters with X-ray luminosity $\geq 5\times 10^{44}$ erg/s and redshift $0.2\leq z\leq 0.32$. Using several methods to characterize cluster substructures, namely the power ratios, centroid shift and X-ray brightness concentration parameter, we show that clusters with and without radio halo can be quantitatively differentiated in terms of their dynamical properties. In particular, we confirm that radio halos are associated to dynamically disturbed clusters and cluster without radio halo are more ``relaxed'', with only a couple of exceptions where a disturbed cluster does not exhibit a halo.

\end{abstract}



\section{Introduction}

Radio and X-ray observations of galaxy clusters prove that thermal and non-thermal components coexist in the intracluster medium (ICM). While X-ray observations reveal thermal emission from diffuse hot gas, radio observations of an increasing number of massive galaxy clusters unveil the presence of ultra-relativistic particles and magnetic fields through the detection of diffuse, giant Mpc-scale synchrotron {\it radio halos} (RHs) and {\it radio relics} (\eg Ferrari et al. 2008; Cassano 2009 for review). 
RHs are the most spectacular evidence of non-thermal components in the ICM. They are giant radio sources located in the cluster central regions, with spatial extent similar to that of the hot ICM.

There is collective evidence in the literature that RHs are found in clusters with significant substructure in the X-ray images, as well as complex gas temperature distribution, which are signatures of cluster mergers (\eg Schuecker et al. 2001; Govoni et a. 2004; Markevitch \& Vikhlinin 2001; Venturi et al. 2008). In particular, in a seminal paper Buote (2001) provided the first quantitative comparison of the dynamical states of clusters with RH discovering a correlation between the RH luminosity at 1.4 GHz and the magnitude of the dipole power ratio $P_1/P_0$.
The RH-merger connection suggests that the gravitational process of cluster formation may provide the energy to generate the non-thermal components in clusters through the acceleration of high-energy particles via shocks and turbulence (\eg Sarazin 2004; Brunetti et al. 2009). The discovery of RHs with very steep spectrum supports the scenario of particle re-acceleration by merger-driven turbulence (\eg Brunetti et al. 2008). 

Recently, deep radio observations of a complete sample of galaxy clusters have been carried out as part of the Giant Metrewave Radio Telescope ({\it GMRT}) RH Survey (Venturi et al. 2007; 2008). These observations confirmed that diffuse cluster-scale radio emission is not ubiquitous in clusters: only ~30\% of the X-ray luminous ($L_X(0.1-2.4\,\mathrm{keV})\geq5\times10^{44}$ erg/s) clusters host a RH. Most importantly, these observations allow to separate RH clusters from clusters without RH, showing a bimodal distribution of these clusters in the 1.4 GHz radio power ($P_{1.4}$) versus X-ray luminosity ($L_X$) diagram (Brunetti et al. 2007): RHs trace the well known correlation between $P_{1.4}$ and $L_X$, while the upper limits to the radio luminosity of clusters with no-RH lie
about one order of magnitude below that correlation (see \eg Fig.4 in Brunetti et al. 2007). The reason for this separation is expected to lie in the mechanism responsible for the origin of radio emitting electrons (Brunetti et al. 2009 and references therein). 
In this Letter, we will show for the first time that clusters with RH and clusters without RH can be quantitatively differentiated also according to their dynamical status. We will use archival {\it Chandra} data of clusters in the {\it GMRT} RH Survey and characterize cluster substructure in the X-ray images adopting different methods.

A $\Lambda$CDM cosmology ($H_{o}=70\,\rm km\,\rm s^{-1}\,\rm Mpc^{-1}$, $\Omega_{m}=0.3$, $\Omega_{\Lambda}=0.7$) is adopted.

\section{The sample and data preparation}
\label{sec:sample}

The {\it GMRT} RH Survey (Venturi et al. 2007, 2008) is a deep, pointed radio survey of clusters selected from the ROSAT--ESO Flux Limited X--ray (REFLEX; B\"oringher et al. 2004) and extended ROSAT Brightest Cluster Sample (eBCS; Ebeling et al. 1998, 2000) catalogs. These two catalogs have almost the same flux limit in the $0.1-2.4$ keV band ($\gtsim 3\cdot 10^{-12} \rm{erg\,s^{-1}\,cm^{-2}}$) and their combination yields an homogeneous, flux-limited sample of clusters.
The {\it GMRT} sample consists of 50 galaxy clusters with z=$0.2-0.4$, X-ray luminosity $L_X>5\times 10^{44}$ erg/s and declination $-30^{\circ}\leq\delta\leq60^{\circ}$. 34 clusters in the sample had no high sensitivity radio information and were observed with the {\it GMRT} at 610 MHz. With the above selection criteria the sample is X-ray luminosity limited up to $z\simeq 0.25$ and X-ray flux limited at higher redshift\footnote{This implies a minimum $L_X\sim 10^{45}$ erg/s at the highest redshift of the sample.}(see Fig. 1 and 2 in Cassano et al. 2008).  
Recently, we have undertaken an extension of the {\it GMRT} RH survey by considering all clusters in the REFLEX and eBCs catalogs with $\delta> 60^{\circ}$ and with the same $z$ and $L_X$ selection. This extension leads to a sample of 67 galaxy clusters which we refer to as the extended {\it GMRT} cluster sample. While the radio campaign is ongoing, data for 3 clusters of the extended sample are available and are considered in the present paper.

For all clusters in the extended {\it GMRT} sample with the radio data at hand, we searched in the {\it Chandra} archive and found information for 
a sub-sample of 35 galaxy clusters. We also required the clusters to have at least 2000 ACIS-S or ACIS-I counts in the 0.5-2 keV band inside an aperture of 500 kpc (see below) in order to produce images sufficient to study the cluster morphological properties. Three of the 35 clusters do not match the requirement and thus are not included in our analysis. Clusters without radio or X-ray data are unlikely to be selected by having/not having a RH or being/not being a merger, so the fact that they are omitted should not affect our conclusions. 
Our final sample consists of 32 galaxy clusters with z=$0.2-0.4$, $L_X>5\times 10^{44}$ erg/s all with radio ({\it GMRT} and Very Large Array, {\it VLA}) and X-ray ({\it Chandra}) data. Only three clusters have $z>0.32$. Indeed it has been shown that for $z>0.32$ the {\it GMRT} sample is incomplete (Cassano et al. 2008). We will thus present results obtained by including and excluding the three clusters at $z>0.32$.

All X-ray images have been produced in a standard manner using CIAO 4.1.2 (with calibration files from the CALDB 4.1.3) in the 0.5-2 keV band. 
They have been renormalized by the exposure maps to maintain unit of counts per pixel and preserve a proper application of the routines for the estimates of the X-ray morphology
parameters. We visually inspected each image to remove point sources and any residual
with low exposure around the CCD gaps.

In this Letter we study the cluster substructure on the RH scale analyzing the surface brightness inside an aperture radius of 500 kpc, since we are interested in the cluster dynamical properties on the scales where the energy is most likely dissipated. Indeed, studies showed a point-to-point correlation between the radio and X-ray brightness (\eg Govoni et al. 2001) and between the RH properties and cluster properties (mass, velocity dispersion) calculated within the halo region (Cassano et al. 2007). The choice of 500 kpc provides a first natural approach, and has also the advantage that it allows us to sample both lower and higher redshift clusters in the sample with adequate sensitivity. Furthermore, we expect a small variation (of about $\sim 1.5$) in $R_{\Delta}$ (the radius defined as that enclosing a region with an overdensity $\Delta=200, 500$, etc. with respect to the critical density at the cluster redshift) among clusters in our sample, because they are characterized by very similar X-ray luminosity and redshift. This implies that our results should not significantly change by considering a radius that takes into account the variation of the cluster thermal properties. We will explore the morphological estimators based on different aperture size in a forthcoming paper (Cassano et al., in prep.).

\section{Morphological estimators}
\label{sec:morph}

The superb angular resolution of {\it Chandra} allows to discriminate between mergers and relaxed clusters even by simple visual inspection. To provide a more quantitative measure of the degree of the cluster disturbance, we use three methods: power ratios 
(\eg Buote \& Tsai 1995; Jeltema et al. 2005; Ventimiglia et al. 2008; 
B\"ohringer et al. 2010); the emission centroid shift (\eg Mohr
et al. 1993; Poole et al. 2006, O'Hara et al. 2006; Ventimiglia et al. 2008, Maughan et al. 2008, 
B\"ohringer et al. 2010) and the surface brightness concentration parameter
(\eg Santos et al 2008). In the following, we briefly describe these methods.

\begin{figure}[t]
\includegraphics[width=0.4\textwidth]
{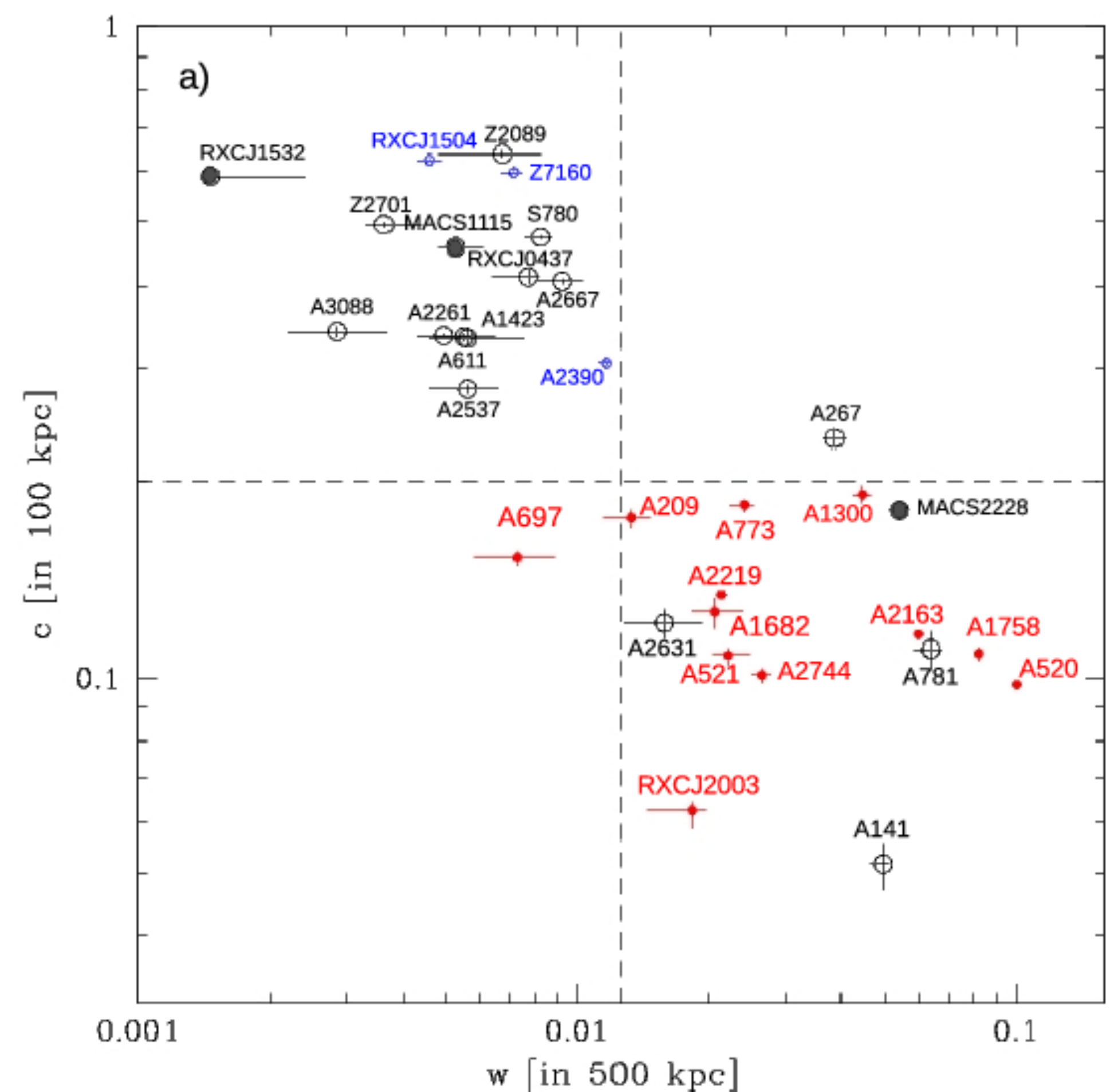}
\includegraphics[width=0.41\textwidth]
{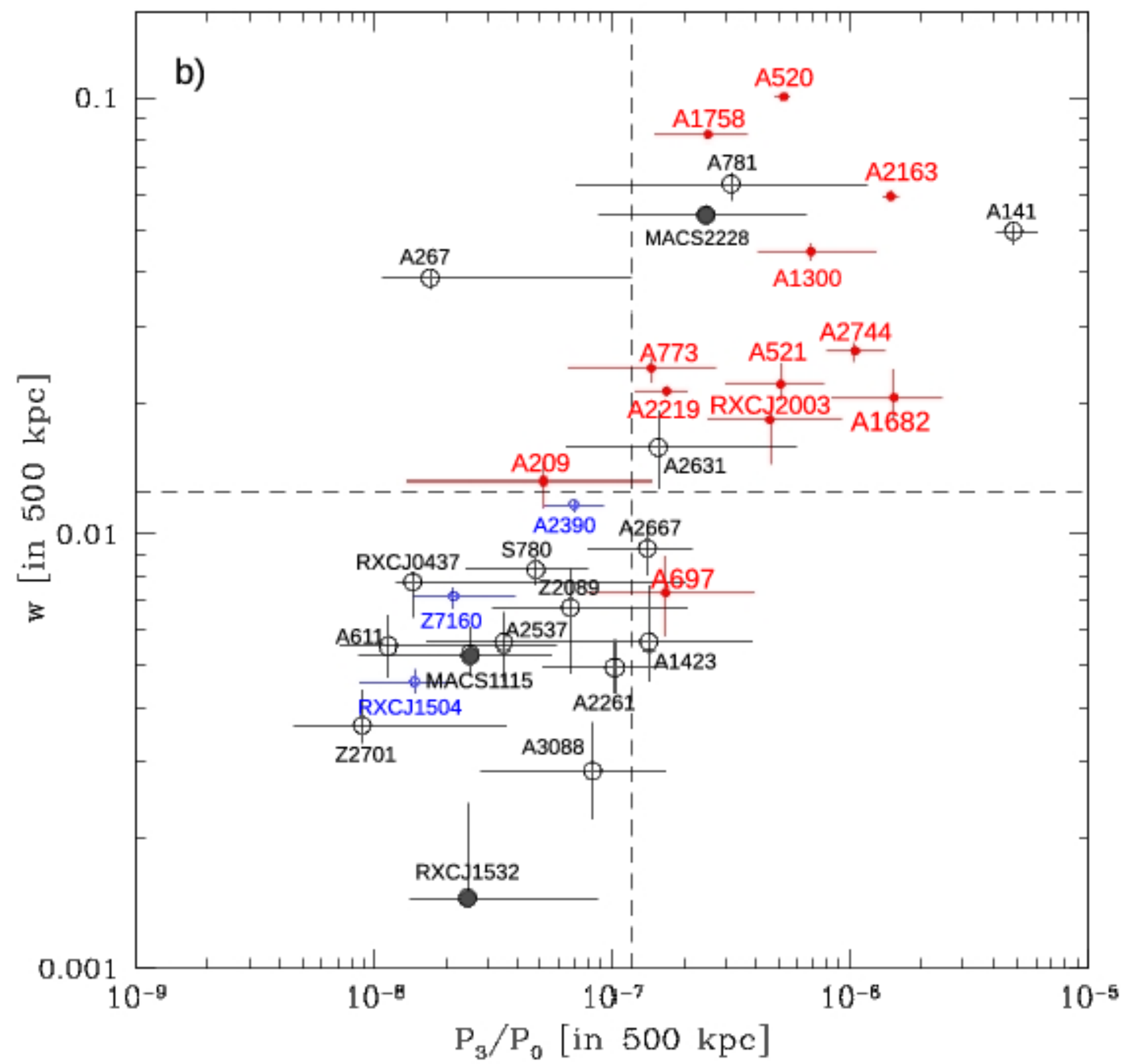}
\vskip0.17cm
\includegraphics[width=0.41\textwidth]
{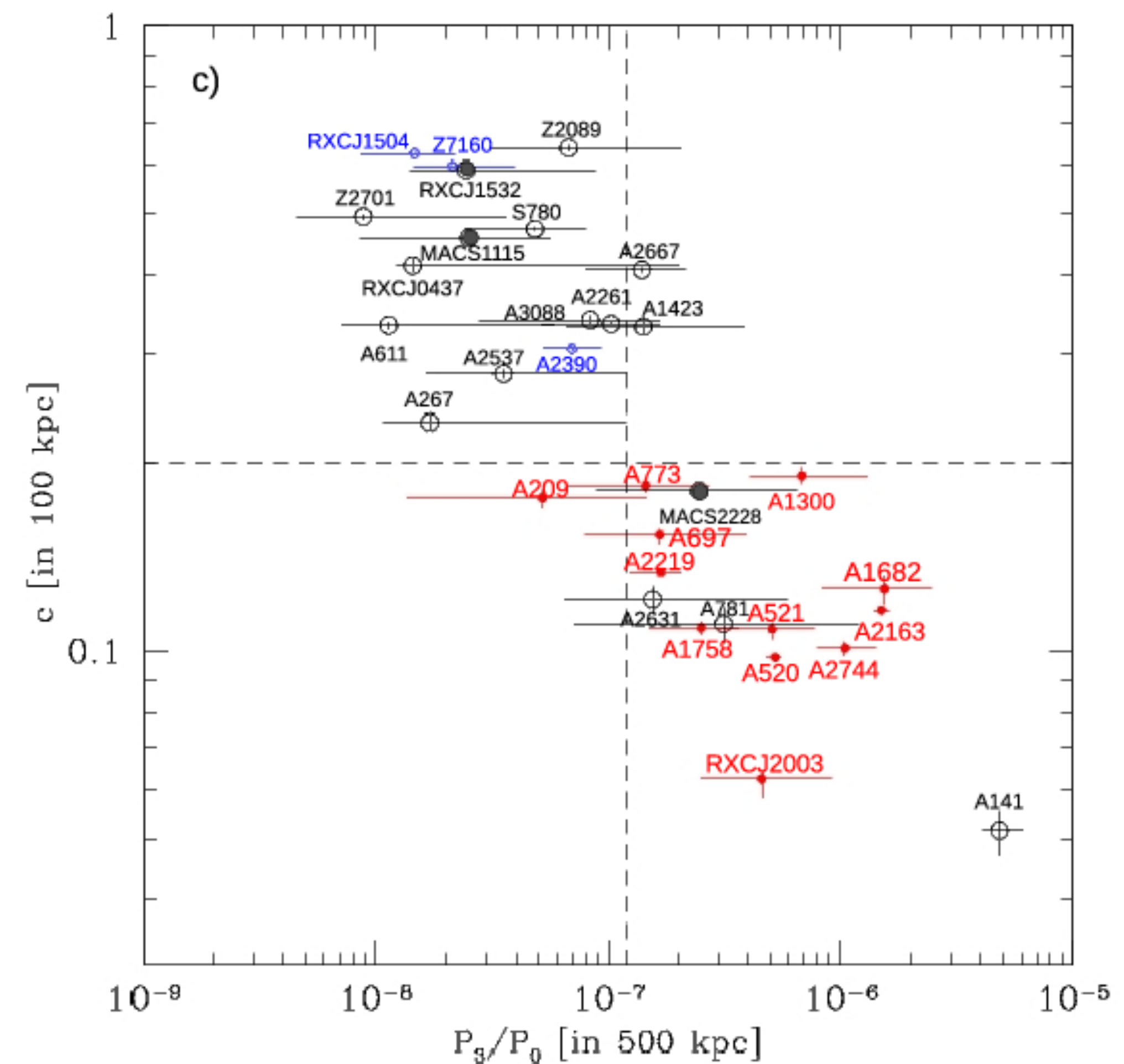}
\caption[]{{\it a):} concentration parameter $c$ vs. centroid shift $w$ (see text for details); {\it b):} $w$ vs. the power ratio $P_3/P_0$; {\it c)} $P_3/P_0$ vs. $c$. 
Symbols are: RH (red filled dots), no-RH (black open dots), mini-halos (blue open dots), $z>0.32$ (gray filled dots). Vertical and horizontal dashed lines are: $c=0.2$, $w=0.012$ and $P_3/P_0=1.2\times 10^{-7}$.} 
\label{fig:substructures}
\end{figure}


\vskip 0.3cm
\noindent
{\it i)} The power ratio method is motivated by the idea that the X-ray surface brightness could represent
the projected mass distribution of the cluster. The power ratio is a multipole decomposition of
the two-dimensional, projected mass distribution inside a given aperture $R_{ap}$. Power ratios 
are usually defined as:

\begin{equation}
P_0=[a_0\,ln(R_{ap})]
\label{Eq:P0}
\end{equation}

\noindent where $a_0=S(<R_{ap})$ is the total intensity inside the aperture radius, and

\begin{equation}
P_m=\frac{1}{2m^2R_{ap}^{2m}}(a_m^2+b_m^2)
\label{Eq:Pm}
\end{equation}

\noindent where the moments $a_m$ and $b_m$ are given by:

\begin{equation}
a_m(R)=\int_{R'\leq R_{ap}} S(x')(R')cos(m\phi')d^2x'
\label{Eq:am}
\end{equation}

and 

\begin{equation}
b_m(R)=\int_{R'\leq R_{ap}} S(x')(R')sin(m\phi')d^2x'
\label{Eq:am}
\end{equation}

\noindent where $S(x)$ is the X-ray surface brightness. 
Here we will make use only of the $P_3/P_0$, that is the lowest power ratio moment providing a clear substructure measure (\eg B\"ohringer et al. 2010).

\vskip 0.3cm
\noindent {\it ii)} Poole et al. (2006) used numerical simulation of cluster mergers and found that the centroid shift method was very sensitive to the dynamical state of the cluster. Following the method of Poole et al. (2006) and Maughan et al. (2008), the centroid shift is computed in a series of circular apertures centered on the cluster X-ray peak. 
The radius of the apertures was decreased in steps of 5\% from $R_{ap}=500$ kpc to $0.05 R_{ap}$, and the centroid shift, $w$, was defined as the standard deviation of the projected separation between the peak and the centroid in unit of $R_{ap}$, as:

\begin{equation}
w=\Big[\frac{1}{N-1}\sum (\Delta_i-\langle \Delta \rangle)^2\Big]^{1/2}\times \frac{1}{R_{ap}},
\label{Eq:w}
\end{equation}

\noindent where $\Delta_i$ is the distance between the X-ray peak and the centroid of the {\it i}th aperture.
\vskip 0.3cm
\noindent {\it iii)} Following Santos et al. (2008), we made use of the concentration parameter, $c$, defined as the ratio of the peak over the ambient surface brightness, $S$, as:

\begin{equation}
c=\frac{S(r<100\,\mathrm{kpc})}{S(<500\,\mathrm{kpc})}
\label{Eq:c}
\end{equation}

\noindent The concentration parameter has been used in literature for a first identification of cool core clusters in those cases where a spatially resolved spectroscopic analysis was not possible (\eg in the case of high redshift clusters; Santos et al. 2008). 
We use the concentration parameter to differentiate galaxy clusters with a compact core (\ie core not disrupted from a recent merger event) from cluster with a spread distribution of gas in the core (\ie core disturbed from a recent merger episode). 

It is important to note that among the presented methods, the power ratio and the centroid shift methods are less sensitive to the presence of substructures (and thus mergers) along the line of sight, while the concentration parameter is in principle not affected by these projection effects.

\section{Results}
\label{sec:results}

We show the results of the substructure analysis in Fig.\ref{fig:substructures}. In Fig.\ref{fig:substructures}a) we report the distribution of the 32 clusters in the ($c,w$) plane.
There is a clear anti-correlation between the two parameters. 
Most importantly, RH clusters (red filled dots) can be well separated from clusters without RH (black open dots) and clusters with mini-halos\footnote{Radio mini-halos are diffuse synchrotron emission on smaller scales (\eg 200-500 kpc) extending around powerful radio galaxies at the center of some cool core clusters (\eg Ferrari et al. 2008).} (blue open dots). In particular, as a reference case, if we consider the median value of each parameter, $w=0.012$ and $c=0.2$ (horizontal and vertical lines), the sample can be well separated between RH and no-RH clusters: no cluster with RH is found in the upper region ($w<0.012$ and $c>0.2$), while the fraction of clusters with RH in the lower region ($w>0.012$ and $c<0.2$) is 73-78 \% (including or excluding the cluster at $z>0.32$, respectively). This confirms the hypothesis that RH are located in dynamically disturbed systems. We note also that clusters with mini-halos lie in the upper region ($w<0.012$ and $c>0.2$), supporting the connection between radio mini-halos and cluster cool cores (\eg Gitti et al. 2002).

In Fig.\ref{fig:substructures}b) we report the distribution of the 32 clusters in the ($w,P_3/P_0$) plane. We find a clear correlation between the two parameters.
A similar trend was recently found also by B\"ohringer et al. (2010). Most importantly, we find that all RH clusters (the color code is the same as above) are located in the region of higher values of the parameters $w$ and $P_3/P_0$. The position of clusters with mini-halos is consistent with being in the more relaxed systems. In this plane the horizontal and vertical lines ($w=0.012$ and $P_3/P_0=1.2\times 10^{-7}$) are also the medians of each parameter.

For completeness, in Fig.~\ref{fig:substructures}c) we report the distribution of clusters in the ($c,P_3/P_0$) plane. There is again a clear separation between RH and no-RH clusters, with RH clusters located in the region of dynamically disturbed systems, with higher values of $P_3/P_0$ and lower values of $c$. We note that $c$ is the parameter that provides the best separation between RH and no-RH clusters, indeed no RH is found in clusters with $c>0.2$. 

All diagrams provide strong evidence that RHs form in dynamically disturbed clusters, while clusters with no evidence of diffuse synchrotron emission on Mpc scales are more relaxed systems. To test quantitatively this result, we run Monte Carlo simulations in the ($w,c$) plane. We randomly distribute the 12 RH clusters among the 32 clusters of the sample, and count the number of RHs falling in the upper-left quadrant of Fig.\ref{fig:substructures}a) (those with $w<0.012$ and $c>0.2$). We repeat the procedure $10^5$ times and find that only in 3-4 cases no RH is found in the upper-left quadrant; this allows us to conclude that the observed distribution differs from a random (\ie independent of cluster dynamics) distribution at more than $4\sigma$.
This proves that our result is statistically significant and shows,  for the first time, that the separation between RH and no-RH clusters has a corresponding separation in terms of dynamical properties of host clusters. We note that there are 4 outliers in Fig.\ref{fig:substructures} (3 if we do not consider the cluster at $z>0.32$), \ie clusters that are dynamically disturbed but that do not host a RH. These clusters deserve further investigation. However, their presence in the region of RHs is not surprising in the framework of models that explain the cluster-scale synchrotron emission with merger-driven turbulence and shocks. These models predict that RH should be maintained for a typical lifetime of $\sim1$ Gyr (see Brunetti et al. 2009) which is of the same order as the merger time-scale (during which the cluster would appear disturbed) implying that a large fraction of massive and merging clusters should host a RH. 
Most importantly, turbulent re-acceleration models predict a cut-off in the spectra of RHs at the frequency $\nu_c$ that is determined by the fraction of turbulent energy converted into electron re-acceleration. The cut-off makes the observations of RHs difficult at $\nu>\nu_c$. 
In disturbed clusters with relatively smaller masses ($M_v\ltsim 2\times 10^{15}\,M_{\odot}$) and at higher redshifts ($z\gtsim 0.4-0.5$) the cutoff frequency can be lower (Cassano et al. 2010, Cassano 2010).
In line with this scenario, 3 out of the 4 outliers have X-ray luminosity at the lower boundary of our selection, $L_X\ltsim 8\times 10^{44}$ erg/sec ($M\ltsim 1.9\times 10^{15}\,M_{\odot}$), and the other is the highest redshift cluster of the sample ($z\simeq0.42$). They may still have RHs, but need lower frequency observations to detect them.

\section{Conclusions}
\label{sec:conclu}

We used a statistical sample of 32 galaxy clusters with $L_X(0.1-2.4\,\mathrm{keV})\geq5\times10^{44}$ erg/s with radio ({\it GMRT} and/or {\it VLA}) and X-ray ({\it Chandra}) observations, to test the merger-RH paradigm in galaxy clusters by relating the dynamical state as seen from X-rays to the presence of a halo.  We adopted three main methods of X-ray substructure characterization: the power ratio, $P_3/P_0$ (\eg Buote \& Tsai 1995; B\"ohringer et al. 2010), the centroid shift, $w$ (\eg Mohr et al. 1993) and the X-ray brightness concentration parameter, $c$ (\eg Santos et al. 2008). We studied the distributions of clusters in the $w-c$, $P_3/P_0-w$ and $P_3/P_0-c$ diagrams (see Fig.\ref{fig:substructures}). As expected, we found anti correlation between $w$ and $c$ and between $c$ and $P_3/P_0$ (clusters with the most compact cores are also less disturbed) and a correlation between $w$ and $P_3/P_0$.
RH and no-RH clusters are clearly separated in all three diagnostic diagrams, with RHs located in more disturbed systems. In particular, the median value of each parameter ($w\simeq0.012$, $c\simeq0.2$ and $P_3/P_0\simeq1.2\times10^{-7}$) splits the sample in RH and no-RH clusters. We find no RH cluster in the regions selected by $w<0.012$, $c>0.2$ and $P_3/P_0<1.2\times10^{-7}$ (also, no RH is found in the region constrained just by $c>0.2$), while the fraction of RHs increases to $\sim 73-78\%$ in the regions selected by $w>0.012$, $c<0.2$ and $P_3/P_0>1.2\times10^{-7}$.  By means of Monte Carlo simulations we showed that the probability to get such segregation between RH and no-RH clusters by chance is of the order of $3-4\times 10^{-5} $. This established for the first time in a statistical manner the connection between RH and cluster mergers. We note also that radio mini-halos are located in relaxed clusters (characterized by high values of $c$) supporting the connection between radio mini-halos and cluster cool cores.

\section{Acknowledgments}
This work is partially supported by INAF under grants PRIN-INAF2007 and PRIN-INAF2008 and by ASI-INAF under grant I/088/06/0. Support for this work is provided by the grant ARO-11017X issued by the {\it Chandra} X-ray Observatory Center.  RC and GB thank the Harvard-Smithsonian Center for Astrophysics for its hospitality. We thank the referee for useful comments.

\end{document}